# CHAINGE: A Blockchain Solution to Automate Payment Detail Updates to Subscription Services

David Buckley*, Gueltoum Bendiab*, Stavros Shiaeles*, Nick Savage *, Nicholas Kolokotronis[†]
*Cyber Security Research Group, University of Portsmouth, PO1 2UP, Portsmouth, UK
UP839643@myport.ac.uk, gueltoum.bendiab@port.ac.uk, sshiaeles@ieee.org, nick.savage@port.ac.uk
[†]Department of Informatics and Telecommunications, University of Peloponnese
22131 Tripolis, Greece. nkolok@uop.gr

*Abstract*—The rise of the subscription-based business model has led to a corresponding increase in the number of subscriptions where a customer needs to manage their payments. This management of payments for multiple subscriptions has become a very complicated and insecure task for customers; especially when it comes to renewing payment details when the card is lost, stolen, or expires. In addition, this, mostly manual, process is vulnerable to human error, digital frauds, and data breaches, according to security reports. Thus, in this paper, we propose a novel approach to automate, manage and simplify the Financial Supply Chain involved in the process of updating and managing payments to user' subscriptions. This is done by utilising the Hyperledger Sawtooth blockchain framework, that allows a consumer to enter their payment card details in a central digital wallet and link their subscriptions to their cards. The card being updated triggers an event on the blockchain, which allow for the payment details to be updated on subscription systems automatically. The verification tests performed on the prototype of the proposed system shows that its current implementation has been securely achieved.

## I. INTRODUCTION

In recent years, it has been seen that the subscription-based business mode as proven itself to be a major success [1]. This model has increasingly become the common way to buy goods and services online, and the growth of that model is not slowing down [1], especially, with the advances made in the emerging area of cloud computing and the rise of Software as a Service (SaaS) applications. Market analyses affirm that subscription as a service has reached significant capacity and proven to be extremely successful, as more and more customers prefer to rent the software rather than buy it [2]. According to McKinsey & Company [3], in the past five years, the online subscription market has grown by more than 100% percent every year, with large sellers generating close to $3B in sales in 2016, up from $57.0 M in 2011. Companies like Netflix, Spotify, Amazon, and Hulu clearly dominate the entertainment subscription segment, and analysts predict that Netflix will have numbers of subscribers by 2028 that equal more than the entire population of the United States [4]. This success is mainly associated with the increasing growth of digital payments and e-wallets [5].

Digital payments platforms such as PayPal, Stripe and WePay have made it possible to set up and process recurring payments without the massive investment of workforce that was once required. It is now more common than ever for consumers to be paying for multiple Subscription Services (SS) at once with either a credit or debit card [5], [6]. However, managing payments for multiple subscription services has become a complicated and laborious task for customers; especially when it comes to renewing payment details [6]. Moreover, if a consumer's card is lost, stolen, or has expired, the customer must go through a manual process in order to not only obtain a new card, but update their payment method on all of their subscription-based services. This process can take the form of multiple steps with multiple transactions needed to be made with each service that the user subscribes to. In addition, this manual process is vulnerable to human error, digital frauds, and data breaches [7]. According to RiskBased Security [8], data breaches hold a large threat for the subscription-based business model, with over 37 billion known records being exposed in 2020.

In this paper, we present a novel system that utilises the Sawtooth blockchain to automate, secure and simplify the management of the Financial Supply Chain (FSC) in- volved in the process of updating and managing payments to subscription-based services. The objective of the proposed system is to provide a central location where users can update all of their payment details for the SSs they pay for. This will be achieved by using an innovative digital wallet solution that will securely and privately hold user credit cards and IDs, accessible only by those who have access to the login credentials. Users can automatically add payment (debit/credit) cards to their digital wallets and link paid services to the cards that they have added, allowing them to see which services they pay for with which card, along with the total outgoings for each card. They can also update their card details from within their digital wallets, which will then push encrypted data onto the blockchain, in-turn informing the linked services of the updated card of the change in payment details. A proof-of-concept prototype of the proposed system has been implemented to prove its effectiveness and reliability. The rest of the paper is structured as follow. Section II reviews the existing methods and representative systems and platforms proposed in this area. Section III is devoted to the description of the proposed solution. Section IV summarises and analysis the obtained testing-results. Finally, Section V concludes the paper, proposes some future work and open issues.

## II. RELATED WORK

The development of automatic, flexible, and secure sub- scription payment has been a priority area for research and development as it has a direct influence on the acceptance and promotion of the subscription-based business model [9], [10]. For instance, authors in [9] proposed a micro-transaction system to provide an efficient, secure and flexible subscription- based solution to e-commerce for digital goods and services. This solution enhanced the user's privacy and authentication aspects of the micro-transactions by integrating a Role- Based Access Control (RBAC) system based on the Public Key Infrastructure (PKI). With the popularity of Bitcoin and blockchain technology, there have been many start-ups looking to replicate the success of Bitcoin for a wide variety of use cases [11]. This technology seems to be a perfect enabler of new levels of collaboration among supply chain actors, especially in the SCF space, by establishing trust between unknown parties in a decentralized network without the need of a third party [10]. Study in [12] suggests that blockchain technology could help speed up and reduce costs of SCF solutions. In addition, this technology can prevent security issues such as 'double spending' whilst improving security and transparency of the whole financial supply chain.

Recently, blockchain-based SSs are gaining popularity as they make sensitive data secure and reliable while protecting it and defining several access levels. A secure subscription protocol is proposed in [13] using blockchain technology. This protocol enables the customers to subscribe to the Cloud Service Providers (CSPs) using fixed subscription and Pay-as- you-go automatic subscription payment models, without the need of a trustful centralised third party (e.g., banks) [14]. This system was implemented by using smart contracts and digital wallets in the Ethereum network. However, it does not address the issue of managing payments for multiple SSs by the customer. Another recent work in [15] proposed a blockchain-based fair payment framework for the payment of subscription-based cloud outsourcing services, including storage and computation. Performance analysis showed that the proposed solution is very efficient in terms of the number of involved transactions and computation cost. In addition, several crypto subscription payment platforms have recently emerged such as such as Groundhog [16], Amberpay [17], Arf [18] and 8Pay [19]. these platforms offer advantages to both SSs and customers. 8Pay [19] is the best platform, and it is a fully decentralized, crypto payment solution that lets SSs manage recurring payments automatically. This platform helps customers to automatically pay for their subscriptions (Netflix, Hulu, etc.) with their favoured digital tokens. The platform does not require personal data or digital identities, customers can send and receive tokens directly from and to their personal wallets [19]. The main problem for this type of solution is that most of the current subsection services refuse payments in cryptocurrency [19]. To the best of our knowledge, there does not exist an application or a service that automates the process of updating the payment methods of the SSs.

## III. PROPOSED SOLUTION

This section presents the proposed system to automate, secure and simplify the management of the Financial Supply Chain (FSC) involved in the process of updating and managing payments to subscription-based services. Using this approach, users can safely update their credit cards details for all of their inscription services in a controlled location, instead of updating them on a per-service basis. This will also bring benefits to the subscription services, as it reduces the risk of missed payments due to customers not updating their payment methods, increasing profits and turnover. The proposed model is presented in Fig. 1.

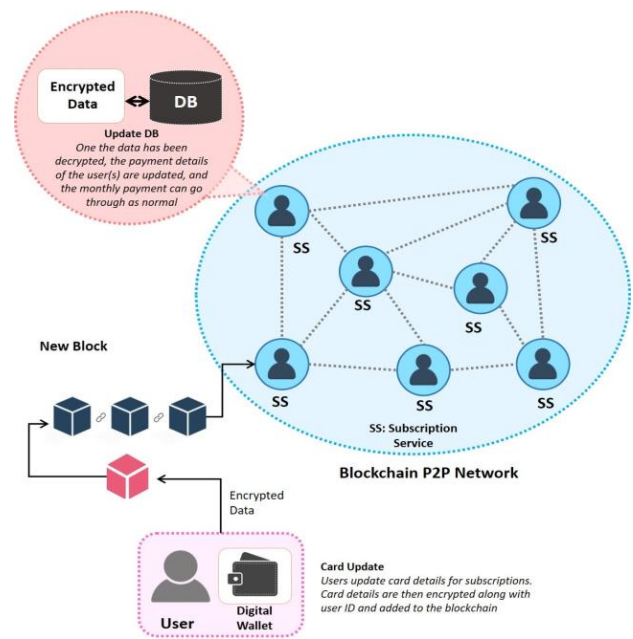

Fig. 1. High-level architecture of the CHAINGE system

As shown in Fig. 1, the core aspects of the proposed system are the blockchain network that incorporates a number of subscription services, and the innovative digital wallet application that enables individuals to store, manage and update their cards and payment details in a central location. Each user in the system has access to its own "digital wallet", where their cards and subscriptions information is stored. The data structure in the wallet allows for multiple cards, with multiple subscriptions to be added, where each card details are only accessible by the user and the linked subscription services. In order to enhance the digital wallet security, parts of the wallet will be encrypted with the user's public key such as sensitive information related to debit/credit cards. In addition, all information exchanged between the user and the subscription services will be encrypted. Users can update cards details from their digital wallet, which will automatically push encrypted cards details along with the user ID into the blockchain, in-turn informing the linked subscription services of the updated cards details. The subscription services can use their private keys to decrypt and read data from the blocks that will contain the updated cards details of consumers and update their records (subscription service DB) accordingly. The card details will be automatically updated for all the subscription services linked to that credit/debit card.

*A. Blockchain: Hyperledger Sawtooth*

The blockchain P2P network is a core aspect of the pro- posed system. It involves a number of subscription services (subscription providers), which constitute the nodes of the blockchain (see Fig. 1). In this work, the Hyperledger Saw- tooth framework [20] was chosen to implement the blockchain for the following reasons:

- Hyperledger Sawtooth offers a flexible and modular architecture that separates the core system from the ap- plication domain [20], which keeps the system safe and secure.
- Hyperledger Sawtooth supports a variety of robust consensus algorithms, including Practical Byzantine Fault Tolerance (PBFT) and Proof of Elapsed Time (PoET) that offers benefits of low resource utilization and low energy consumption.
- Due to its modular architecture, Hyperledger Sawtooth can specify business rules without having to know the design of the system.
- Support Ethereum Smart Contracts, which enables Saw- tooth to use Ethereum standards like ERC-20 (Ethereum Request for Comments 20) [21] and allows for inter- operability between Sawtooth projects and the Ethereum ecosystem.

As illustrated in Fig. 2, the Sawtooth P2P network incorporates a number of different components including validators, Transaction Processors, Event Subscribers, REST APIs and the consensus engines. Each node in the Sawtooth network runs a single validator, a REST API, a consensus engine, and one or more transaction processors.

- **Validator**: is responsible for validating groups of related transactions, including them into blocks, maintaining consensus with the Sawtooth network, and coordinating communication between clients, transaction processors, and other validators on the network [22].

- **Transaction Processor**: is responsible for validating transactions and updating state based on the rules defined by the associated transaction family. It also handles the data stored on the blockchain, validates that data, correctly formatting it and signing it in order to be sent to the validator [22].

- **Rest API**: is a core component in the Sawtooth network that is used to exchange data between the various com- ponents of the system [22].

- **Consensus Engine**: provides consensus-specific functionality for a Sawtooth node. The consensus engine runs as a separate process on the node (see Fig. 2) and communicates with the validator through the consensus API [22]. The Sawtooth consensus engine allows consensus algorithms to be changed on-chain. The initial consensus algorithm is set through on the genesis block but can be changed with the settings transaction processor [23].

- **Event Subscriber**: subscribes to core Sawtooth specific events that occur on the blockchain, such as a new block being committed [22]. In our work, this component is used to represent subscription services, where each subscription service will subscribe to events specific to their service. When an event subscriber detects a card change, a custom transaction event is created and used by linked subscription services to update their records (i.e., subscription service DB) in accordance with the new card details.

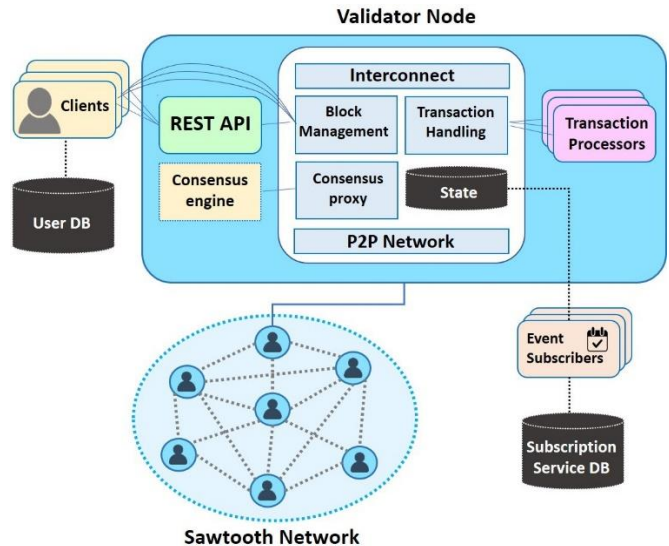

Fig. 2. Hyperledger Sawtooth components diagram [23]

In this work, the client forms the front-end section of the system in the form of a web application; when a user signs in, their information will be stored in a database (i.e., user DB) along with their public key. Clients can add their paid subscription services, linked to the relevant payment card, to the blockchain using the web application.

*B. Digital Wallet*

As highlighted before, each user (or client) in the system has it is own digital wallet, where their debit/credit cards and subscription details are stored. The wallet is composed of an array of cards and linked subscriptions. Each card has a unique identifier (card ID), an alias (a user-friendly name to display the card), and the card-related data (e.g., card number, expiry date, etc.). Each subscription has a subscription type, the user ID linked to the subscription service. This user ID is generated when the user creates an account with that subscription service. Fig. 3 shows how this information will be stored in a user's wallet. For every single wallet, the information is stored in a JSON file.

Fig. 3. Digital wallet data structure

*C. System operations*

The sequence diagram presented in Fig. 4 gives a high-level view of the operations that can be performed by a registered user in the system. As highlighted before, all user-interaction with the system is handled by web applications and APIs. Certain aspects of the sequence diagram have been simplified in order to make understanding easier. For instance, a "Transaction" combines different actions in the system including adding a new card or a new subscription service, updating card details, or removing a card from the wallet. As shown in the diagram, the user can use SSO authentication mechanisms (3rd party provider) in order to login to the system. The data on the blockchain is only accessible if the user has successfully authenticated with the 3rd party provider and has entered the correct private key. The correctness of the user private key is verified using the user public key stored in the user DB. The pair of keys were generated during the first login of the user to the system, with the public key being stored in the user DB and the private key being displayed to the user. Once the user has successful access to the system, he can perform the flowing operations:

- Add cards to the wallet in form of transactions that will be added to the blockchain. The transactions are not immediately added to the blockchain; instead, they are packed into blocks containing many transactions for efficiency reasons. Once a block is successfully added to the blockchain, appropriate user feedback should be given.

- Link subscriptions to a card. Each subscription details will be validated by authenticating with the subscription service; the details are then encrypted and added to the blockchain in form of a transaction.
- Update the payment details of a card, which are then added to the blockchain in a form of transactions. This operation will trigger the event subscriber that will automatically update the card details for all linked subscription services.
- Remove a card from the system. This will also remove all associated subscriptions with that card but will not cancel the subscriptions with the subscription services.
- This action is also added to the blockchain in a form of a transaction.

IV. SYSTEM IMPLEMENTATION & TESTING

This section describes the tests carried out over the proposed system in order to demonstrate its effectiveness and reliability. To this end, we have developed a proof-of-concept prototype that implements the system. The prototype has been built using Docker, allowing for the code to be split into sections, or containers. Thus, the different components of the system have been implemented using the containers presented in Table. I. Each container is able to interface with each other, independent of the operating system it is running on. The implementation is performed using NodeJS and Node Package Manager (npm). The client application is developed using ReactJS, TypeScript and JavaScript. Whereas the system API was developed using ExpressJS.

*A. Implementation limitations*

SSO system was implemented using google and Office365 for this proof of concept. However, more providers can be implemented in order to give users enough variety when signing in. Also, users are only able to link their Spotify accounts, but many more subscription services need to be offered in order to the system to show its true benefits.

*B. System testing against functional requirements*

Several tests were performed on the proof-of-concept prototype in order to determine the feasibility of the proposed solution. For that, a set of seven functional requirements has been proposed to check that the current scope of the proposed system has been achieved. Each high-level functional requirement is present by a use case that provides a simplified oversight to how core functions of the proposed system work. Table. II summarises the proposed functional requirements.

In terms of verification tests, we run a considerable amount of test on the proof-of-concept prototype against the proposed functional requirements. All the verification tests have been successfully passed and all the proposed functional requirements have been met. The user is able to sign on with his Google account (SSO authentication) (#Req1), and only once authenticated and provided the correct private key, he can access the private areas of the system (#Req2). Also, when the public/private key pair is generated, the public key is displayed and automatically stored in the user database, whereas the private is never sent to the server-side via the REST API, it is only used locally in the user digital wallet (#Req3).

The user is also able to add and retrieve data to/from the blockchain, only if he has successfully authenticated (#Req4, #Req7). Sensitive card details are first encrypted before being sent to the blockchain through the REST API. Thus, it would not be easily readable to intruders unless they had the user's private key (#Req4). It is also proven from the tests that the user is able to link their Spotify account to a payment card. However, there is only currently one subscription service (Spotify) that a user is able to do this with (#Req5). This is acceptable for the proof of concept, but in the future, more subscription services will be added to the system and tested. Moreover, when a user updates their card details, the event subscriber successfully updates the records in the database (subscription service) (#req6). Despite this requirement being met, it could be improved.

The event subscriber does not work with any custom application events, only state changes; custom events would mean that the subscriber would not have to parse a whole state change when detecting updates to cards, meaning that subscription events could be set up that are specific to the subscription type. Overall, it can be seen that the majority of the requirements have been met. For an initial proof of concept, this can be deemed acceptable, as its purpose is to demonstrate core functionality which may be adapted or changed in later iterations of the system.

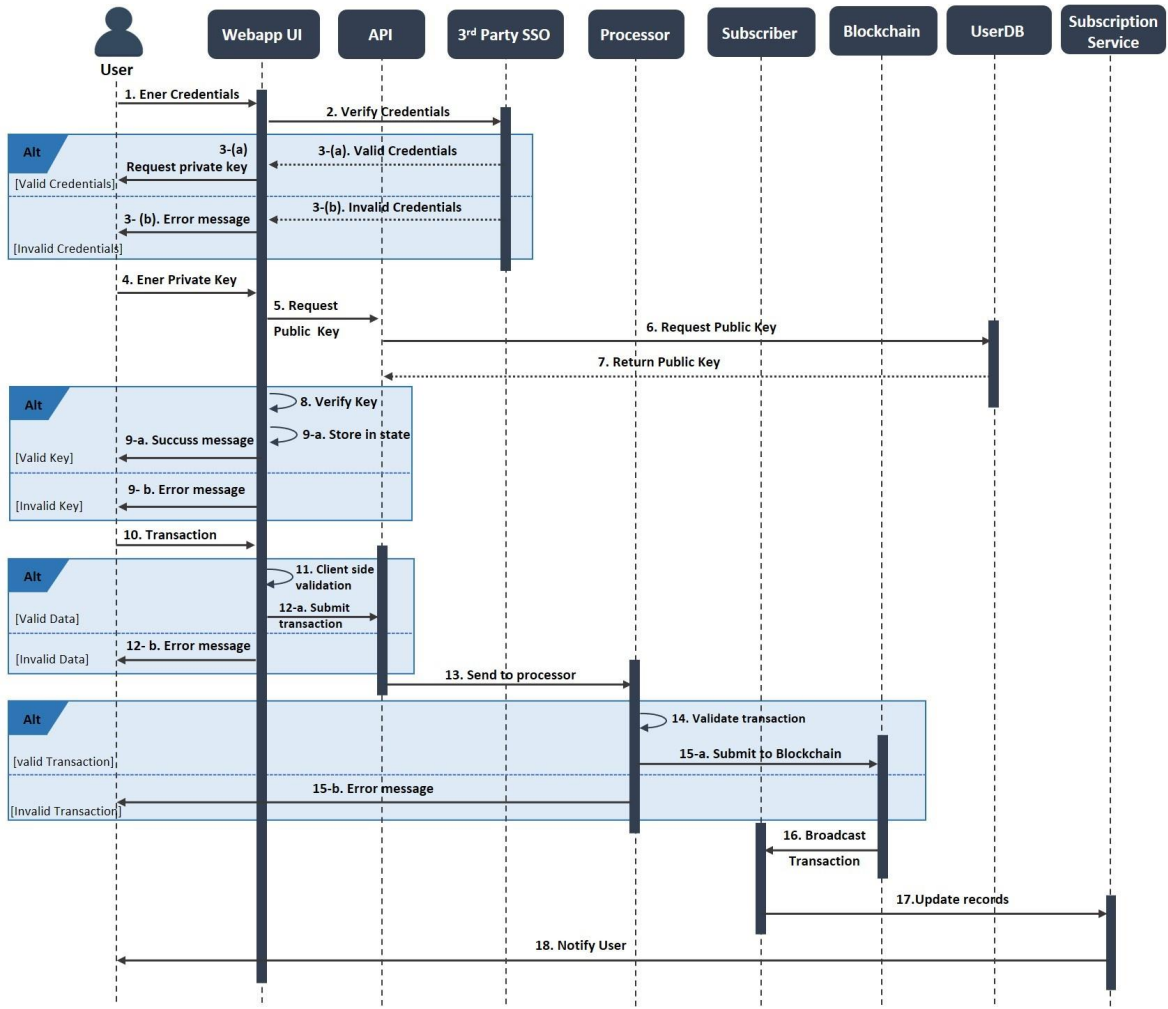

Fig. 4. Sequence diagram of main operations in the CHAINGE system

TABLE I: Containers that implement the system components

| Container | Discerption |
|---|---|
| Sawtoothsubscriptions-js-processor | This container utilises the Sawtooth-SDK and custom code in order to process the various types of transactions described in the previous section. Each transaction type has an associated function, which can make changes to the current state of the block and then publish new blocks. |
| Sawtootsubscriptions-client-js | This container runs the front end of the whole application, as well as the expressJS API, which passes data from the front-end, to the back end to be published as a block on the blockchain. |
| Sawtooth-rest-api | This container is a pre-built image provided by Hyperledger to allow the client to interact natively with the validator with HTTP and JSON standards. |
| Settings-tp | This container is a pre-built image provided by Hyperledger, which runs a single transaction processor dedicated to handling on-chain settings changes such as the consensus algorithm. |
| MongoDB | This container is developed to increase the number of tests when running the project. Thus, when a new account is created, it would generate a public/private key pair and save the public key to the database. This account would then only be valid for the duration of that test or development session. This meant that the same Google account could be used to create a new account each time the system was run |
| . Subscriber | Currently, the subscriber is set up in a way to subscribe only to state changes on the blockchain. However, Sawtooth allows for event subscriptions to happen on specific events. In extended versions of the system, subscribers could be set up in a way to subscribe to only events that involve them; for example, Spotify could subscribe to events that involve card details changing for a user that has a Spotify account. |
| Rethink | This container is a pre-built image which is used to host the database where the updated payment details are added. It is used to simulate the database of a subscription service. |

TABLE II: Functional Requirements

| Requirements | Discerption |
|---|---|
| #Req1 | The system should allow the user to sign into the system using SSO. |
| #Req2 | The user must only be able to view data on the blockchain once successfully authenticated and having provided the private key. |
| #Req3 | The system should assign a unique public and private key for when a user first signs in. |
| #Req4 | The user should be able to add multiple payment cards to the blockchain. |
| #Req5 | The user should be able to link multiple subscriptions to a card and it then be added to the blockchain. |
| #Req6 | The user should be able to update the details of a card. |
| #Req7 | The user should be able to remove card details from the blockchain. |

## V. Conclusion

In this paper, we proposed a Blockchain as a Service (BaaS) solution, called CHAINGE, to automate, manage and simplify the Financial Supply Chain (FSC) involved in the process of updating and managing payments to subscription-based services. In order to check the feasibility of the proposed system, we have developed a proof-of-concept prototype that implements it. Tests performed on the prototype showed that the core functionality of the proposed system has been achieved. However, some improvements are needed in order to add more functionalities. As future work, we intend to conduct additional research and development, to not only improve security and functionality of the current implementation, but to also abstract the idea to work differently and with varying existing infrastructures. In this context, several new ideas will be adopted and used in the CHAINGE system. For instance, payment tokens will be used to improve the system security. This kind of payment will remove the need for a user's full card details to be stored on the blockchain. We also intend to add new functionalities to the system such as cancelling subscriptions and transferring ownership of a subscription to a new card.

## Acknowledgment

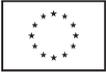 This project has received funding from the European Union's Horizon 2020 research and in- novation programme under grant agreement no. 833673. The work reflects only the authors' view, and the Agency is not responsible for any use that may be made of the information it contains.